\documentclass{aa}
\usepackage{graphicx}
\usepackage{txfonts}
\usepackage{natbib}
\bibpunct{(}{)}{;}{a}{}{,}
\usepackage{longtable}
\usepackage{lscape}
\usepackage{times}
\usepackage{textcomp}
\usepackage{url}

\begin{document}

\title{Thermal and non-thermal emission in the Cygnus X region}

\subtitle{}
\author{W. F.~Xu\inst{1,2}, X. Y.~Gao\inst{2}, J. L.~Han\inst{2}, and
  F. S.~Liu\inst{1}}
\titlerunning{Thermal and non-thermal emission in the Cygnus X region}
\authorrunning{W. F.~Xu et al.}

\offprints{XYG: \email{bearwards@gmail.com}}

\institute{College of Physical Science and Technology, Shenyang Normal
  University, Shenyang 110034, PR China \and National Astronomical
  Observatories, Chinese Academy of Sciences, Jia-20 Datun Road,
  Chaoyang District, Beijing 100012, PR China}

\date{Received; accepted}

\abstract
{Radio continuum observations detect non-thermal synchrotron and
  thermal bremsstrahlung radiation. Separation of the two different
  emission components is crucial to study the properties of diffuse
  interstellar medium. The Cygnus X region is one of the most complex
  areas in the radio sky which contains a number of massive stars and
  \ion {H}{II} regions on the diffuse thermal and non-thermal
  background. More supernova remnants are expected to be discovered.}
{We aim to develop a method which can properly separate the
  non-thermal and thermal radio continuum emission and apply it to the
  Cygnus X region. The result can be used to study the properties of
  different emission components and search for new supernova remnants
  in the complex.}
{Multi-frequency radio continuum data from large-scale surveys are
  used to develop a new component separation method. Spectral analysis
  is done pixel by pixel for the non-thermal synchrotron emission with
  a realistic spectral index distribution and a fixed spectral index
  of $\beta = -2.1$ for the thermal bremsstrahlung emission.}
{With the new method, we separate the non-thermal and thermal
  components of the Cygnus X region at an angular resolution of
  9$\farcm$5. The thermal emission component is found to comprise 75\%
  of the total continuum emission at $\lambda$6\ cm. Thermal diffuse
  emission, rather than the discrete \ion{H}{II} regions, is found to
  be the major contributor to the entire thermal budget. A smooth
  non-thermal emission background of 100~mK\ T$_{b}$ is found. We
  successfully make the large-extent known supernova remnants and the
  \ion{H}{II} regions embedded in the complex standing out, but no new
  large SNRs brighter than $\rm \Sigma_{1GHz} = 3.7 \ \times
  \ 10^{-21}\ W\ m^{-2}\ Hz^{-1}\ sr^{-1}$ are found.}
{}


\maketitle
\section{Introduction}

The Galactic radio emission at decimetre and centimetre wavelengths is
composited by the non-thermal synchrotron radiation generated by the
relativistic electrons spiralling in the Galactic magnetic fields and
the thermal bremsstrahlung (free-free) radiation originated from the
interactions between electrons and ions. Separating the non-thermal
and thermal emission components in the complex areas of the Galaxy
helps to delineate the distribution of the electrons with different
energy spectra, and identify new radio sources buried in the confusing
environments.

To study the properties of interstellar medium and the foreground
emission of the Cosmic Microwave Background, several methods have been
developed to separate the Galactic emission into non-thermal
synchrotron and thermal free-free emission. \citet{Hinshaw07} used the
maximum entropy method for the component separation for the WMAP data,
where the extinction-corrected H$\alpha$ template \citep{Finkbeiner03,
  Bennett03} was used for the derivation of thermal free-free
emission. Using the Effelsberg $\lambda$21\ cm \citep{Reich9021},
$\lambda$11\ cm survey data \citep{Reich9011} together with the Parkes
$\lambda$6\ cm data \citep{Haynes78}, \citet[][here after
  P05]{Paladini05} made the component separation for a thin layer of
the inner Galactic plane ($20\degr \leqslant \ell \leqslant 30\degr,
|b| \leqslant 1\fdg5$) by fitting the observed Galactic emission by
the thermal free-free emission with a fixed brightness temperature
spectral index of $\beta = -2.1$ ($\alpha = \beta + 2$) and the
non-thermal synchrotron emission with a 10$\degr$ resolution spectral
index distribution derived by \citet{Giardino02}. With a similar
method, \citet{Sun11a} (hereafter S11) separated a much larger plane
area of the inner Galaxy ($10\degr \leqslant \ell \leqslant 60\degr,
|b| \leqslant 4\degr$) by using the Effelsberg $\lambda$21\ cm
\citep{Reich9021}, $\lambda$11\ cm \citep{Reich9011}, and the Urumqi
$\lambda$6\ cm data, but with a fixed synchrotron spectral
index. Using the radio recombination line as the tracer of thermal
free-fee emission, \citet{Alves12} separated the non-thermal and
thermal emission component for the region between $20\degr \leqslant
\ell \leqslant 40\degr$ and $|b| \leqslant 4\degr$. They claimed that
the WMAP result greatly overestimates the thermal component in this
area.

All these studies advanced our knowledge about the Galactic
interstellar medium. However, WMAP has a coarse angular resolution of
about $1\degr$, which cannot be used for the studies of small-scale
objects. The synchrotron spectral index distribution map used by P05
has an angular resolution of 10$\degr$ and was derived by assuming
that all the 408~MHz emission is purely non-thermal synchrotron
emission. A fixed synchrotron spectral index assumed by S11 is too
ideal for a practical situation.

With currently available data sets, we plan to develop a new method of
component separation based on the work of P05.  We aim to take a
realistic synchrotron spectral index distribution with a higher
angular resolution and study the properties of the non-thermal
synchrotron and thermal free-free emission. The region we choose is
the Cygnus X region, which is located at about $\ell \sim 80\degr$ in
the Galactic plane. It is one of the most famous and complex star
forming sites in the Galaxy. It was first discovered and presented as
a very extended bright source in the early radio observations
\citep{Piddington52} and later resolved into many individual
components by follow-up observations with higher angular resolutions
and sensitivities \citep[e.g.][]{Wendker91,Taylor96,
  Landecker10}. Although many efforts have been made to study this big
structure, the physical nature of the Cygnus X region remains unclear
and is still under debate. The controversy lies on whether the Cygnus
X region is a structure that the Local Arm is seen end-on
\citep{Wendker91}, or it is a huge coherent and physically bounded
structure along the line of sight \citep[e.g.][]{Knodlseder04,
  Schneider06}. Over two thousand massive stars are located in the
Cygnus X region \citep{Knodlseder00}. Therefore supernova explosions
are expected and many SNRs should exist consequently. Observations
made in the X-ray regime revealed a large bubble around the Cyg OB2
association, which might implies 30-100 supernovae explosions
\citep{Cash80}. However, up to date, only 12 SNRs have been found
($66\degr \leqslant \ell \leqslant 90\degr$, $|b| \leqslant 4\degr$ )
according the Green SNR catalogue \citep{Green09}. It is thus
interesting to search for new SNRs in the Cygnus X region. The very
first step is to remove the strong confusion from the unrelated
thermal emission.

\begin{table*}[!htb]
\centering
\caption{Basic parameters of survey data in this work}
\label{parameter}
{\begin{tabular}{rrrccc} \hline\hline
    \multicolumn{1}{c}{Surveys}  &\multicolumn{1}{c}{Frequency}   &\multicolumn{1}{c}{HPBW} &\multicolumn{1}{c}{ Survey References}  &\multicolumn{1}{c}{T$_{off}$} &\multicolumn{1}{c}{References for T$_{off}$}\\
    &\multicolumn{1}{c}{(MHz)}         &\multicolumn{1}{c}{($\arcmin$)}  &   &\multicolumn{1}{c}{(K)}  &\\
    \hline
    $\lambda$73.5 cm           &408            &51    &\citet{Haslam82}      &2.7        &\citet{PReich04} \\
    Stockert $\lambda$21 cm     &1420           &35.4  &\citet{Reich82}       &2.8        &\citet{Reich88b} \\
    Effelsberg $\lambda$21 cm  &1408           &9.4   &\citet{Reich9021}     &2.8        &\citet{Reich9021}\\
    WMAP $\lambda$1.3 cm       &22800          &52.8  &\citet{Jarosik11}     &$\cdots$   & \\
    \hline
    Urumqi $\lambda$6 cm       &4800           &9.5   &\citet{Xiao11}        &$\cdots$   & \\
    Effelsberg $\lambda$11 cm  &2700           &4.3   &\citet{Fuerst90, Reich9011}       &$\cdots$   & \\
    \hline\hline
\multicolumn{6}{l}{{\bf Note:} T$_{off}$ = T$_{ex}$ + T$_{cmb}$ + T$_{zero}$, see Sect.~3 for detail.}
\end{tabular}}
\end{table*}

In this paper, we develop a method for the non-thermal and thermal
component separation, and search for new SNRs in the Cygnus X
complex. We introduce the data sets we used in Sect.~2 and present the
method and the test results in Sect.~3. We show our separation results
of the Cygnus X region and make discussions in Sect.~4. A summary is
given in Sect.~5.

\section{Data}

\subsection{Data for construction of spectral index distribution of
  the Galactic emission and synchrotron emission}

A proper separation of the thermal and non-thermal emission requires
absolutely calibrated multi-frequency radio continuum
data. Observations made at very high frequencies cannot be used, since
the thermal dust and the anomalous microwave emission originated from
the spinning dust will contribute to the observed radio emission; very
low radio frequency observations also have to be skipped, because the
thermal free-free emission will become optically thick and does not
have the power law spectrum with the spectral index of $\beta \sim
-2.1$. Therefore, data of the 408~MHz survey \citep{Haslam82}, the
Stockert 1420~MHz survey \citep{Reich82}, the Effelsberg 1408~MHz
survey \citep{Reich9021} and the WMAP 22.8~GHz survey
\citep{Jarosik11} are selected in our work to construct a realistic
spectral index distribution for the observed Galactic emission and the
synchrotron radiation at an angular resolution of 1$\degr$. The basic
observational parameters of these surveys are listed in
Table~\ref{parameter}.

\subsubsection{The 408~MHz survey}

The 408~MHz all sky survey was combined by the survey data from the
Jodrell Bank 76-m radio telescope, the Effelsberg 100-m radio
telescope and the Parkes 64-m radio telescope \citep{Haslam82}. The
final released total intensity data has an angular resolution of
51$\arcmin$ and was absolutely calibrated by the 404~MHz survey data
\citep{Toth62}.  The uncertainty of the 408~MHz total intensity data
is better than 10\%.

\subsubsection{The Stockert 1420~MHz survey}

The Stockert 1420~MHz survey covered the northern sky by using the
25-m Stockert radio telescope \citep{Reich82}. The angular resolution
of the survey is 35$\farcm$4. The absolute calibration of the total
intensity data was done by comparison with the horn observations
\citep{Howell66}. The base level error and the uncertainty through the
measurement of a single source are 0.5~K and 5\%, respectively. By
combining this survey with the 408~MHz survey, \citet{Reich88b}
derived the spectral index distribution of the observed Galactic
emission of the northern sky, which is widely accepted and used.

\subsubsection{The Effelberg 1408~MHz survey}
The Effelsberg 1408~MHz survey is a Galactic plane survey covering the
range of $-$3$\degr \leqslant \ell \leqslant 240\degr$ in the Galactic
longitude direction and $|b| \leqslant 4\degr$ in the latitude
direction \citep{Reich9021, Reich97}. The angular resolution of the
survey is 9$\farcm$4. The Stockert 1420~MHz total intensity data was
used to calibrate the 1408~MHz survey. The r.m.s. of the survey is
about 40~mK T$_{b}$.

\subsubsection{The WMAP K-band (22.8~GHz) survey}

The K-band (22.8~GHz) WMAP data is at the lowest frequency among the
five bands taken by the satellite. It contains negligible
contributions of the thermal dust and the anomalous microwave
emission. The angular resolution of the K-band data is
52$\farcm$8. The 7-year released data \citep{Jarosik11} is used in
this study. The recently released 9-year data \citep{Bennett12} is
discussed.

\subsection{Data for high angular resolution component separation}

To achieve a high angular resolution component separation,
observations with better resolutions are needed. The Effelsberg
1408~MHz survey, the Effelbserg $\lambda$11\ cm and the Urumqi
$\lambda$6\ cm survey have an angular resolution of 9$\farcm$4,
4$\farcm$3, and 9$\farcm$5, respectively, and therefore are used for
this purpose. Note, however, that absolute calibration was not done
for the Urumqi $\lambda$6\ cm total intensity data at all, and only
partially done for the Effelsberg $\lambda$11\ cm survey in the Cygnus
X region. The missing large-scale emission component must be restored
in prior to the component separation (see details in Sect.~3).

\subsubsection{The Urumqi $\lambda$6\ cm survey}

The Urumqi $\lambda$6\ cm data of the Cygnus X region was extracted
from a section of the Sino-German $\lambda$6\ cm polarization survey
of the Galactic plane
\citep{Xiao11}\footnote{http://zmtt.bao.ac.cn/6cm}. The observations
were conducted by the Urumqi 25-m radio telescope. The survey covers
the Galactic plane from 10$\degr$ to 230$\degr$ in the Galactic
longitude and $|b| \leqslant 5\degr$ in the latitude, with an angular
resolution of 9$\farcm$5. The uncertainty measured from a
structureless area in the Cygnus X region was about 4~mK T$_{b}$.

\subsubsection{The Effelsberg $\lambda$11\ cm survey}

The Effelsbserg $\lambda$11\ cm survey was conducted by the Effelsberg
100-m radio telescope \citep{Fuerst90, Reich9011}. This survey has an
angular resolution of 4$\farcm$3. The missing large-scale component
was added by comparing with the Stockert $\lambda$11\ cm survey data
\citep{Reif87}, but only for the area of $-2\degr \leqslant \ell
\leqslant 76\degr$, $|b| \leqslant 5\degr$. The r.m.s of the survey
data was about 25~mK T$_{b}$.

\begin{figure}
\centering
\includegraphics[angle=-90,width=0.8\columnwidth]{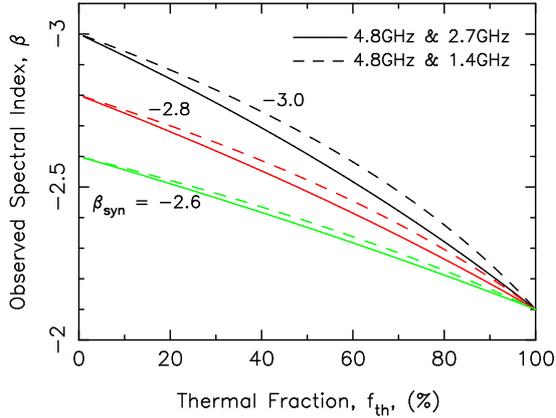}
\caption{Thermal fraction at $\lambda$6\ cm versus the spectral index
  $\beta$ of observed Galactic emission. The spectral index of thermal
  free-free emission is fixed to be $-$2.1, while the spectral index
  of non-thermal synchrotron emission is assumed to be $-$3.0 (black
  lines), $-$2.8 (red lines), and $-$2.6 (green lines),
  respectively. The curves calculated between $\lambda$6\ cm and
  $\lambda$11\ cm is shown by solid line and $\lambda$6\ cm and
  $\lambda$21\ cm by dashed line.}
\label{theoreticalines}
\end{figure}

\section{Method and test}

\subsection{Method}

Our method for the component separation is based on the work of
P05. The observed radio brightness temperature at frequency $\nu$
consists of several contributions \citep{Reich88b}:

\begin{equation}
 T_{obs}(\nu) = T_{gal}(\nu) + T_{cmb} + T_{ex}(\nu) + T_{zero}(\nu),
\end{equation}

\noindent here, $T_{gal}$($\nu$) is the Galactic emission, composed by
the thermal free-free and the non-thermal synchrotron emission,
$T_{cmb}$ is the 2.73~K contribution of the cosmic background,
$T_{ex}$ represents the component of the un-resolved extragalactic
sources, and $T_{zero}$ is the zero level of the data set to be
corrected. The last three contributions comprise T$_{off}$ (T$_{cmb}$
+ T$_{ex}(\nu)$ + T$_{zero}(\nu)$), which should be subtracted first
before the component separation. We got the values of T$_{off}$ for
each survey in literature and present them in Table~\ref{parameter}.

After the T$_{off}$ correction, we have the brightness temperature of
the observed Galactic emission as being:

\begin{equation}
 T_{gal}(\nu) = T_{th}(\nu) + T_{syn}(\nu),
\end{equation}

\noindent here, $T_{th}$($\nu$) and $T_{syn}$($\nu$) are the thermal
free-free and the non-thermal synchrotron radiation observed at a
frequency $\nu$. Both of the two emission contributions have power law
spectra but with different spectral indices. For observations at two
different frequencies, we have:

\begin{equation}
 T_{th}(\nu_1) = {{(\frac{\nu_1}{\nu_2})}^{\beta_{th}}}T_{th}(\nu_2),
\end{equation}

\begin{equation}
 T_{syn}(\nu_1) = {{(\frac{\nu_1}{\nu_2})}^{\beta_{syn}}}T_{syn}(\nu_2),
\end{equation}

\noindent here $\beta_{th}$ is the brightness temperature spectral
index of the thermal free-free emission, while $\beta_{syn}$ is that
for the non-thermal synchrotron emission. As noted by P05, two power
laws added together does not make another power law. However, we can
always find a directly derived spectral index ``$\beta$'' for the
observed brightness temperatures at two bands (see Equation 5). We
show the changes of $\beta$ for wavelength pairs as a function of
thermal emission fraction in Fig.~\ref{theoreticalines}. We emphasize
here that $\beta$ varies with different frequency pairs and with
thermal (non-thermal) fractions.

\begin{equation}
 T_{gal}(\nu_1) = {{(\frac{\nu_1}{\nu_2})}^{\beta}}T_{gal}(\nu_2),
\end{equation}

Combining the Eq.~3 to 5, as done by P05, we easily derived the
thermal fraction at frequency $\nu_1$ as:

\begin{equation}
f_{th_{\nu_1}} = \frac{1 - (\frac{\nu_2}{\nu_1})^{\beta -
    \beta_{syn}}} {1 - (\frac{\nu_2}{\nu_1})^{\beta_{th} -
    \beta_{syn}}},
\end{equation}

\begin{figure*}
\centering
\includegraphics[angle=-90,width=0.3\textwidth]{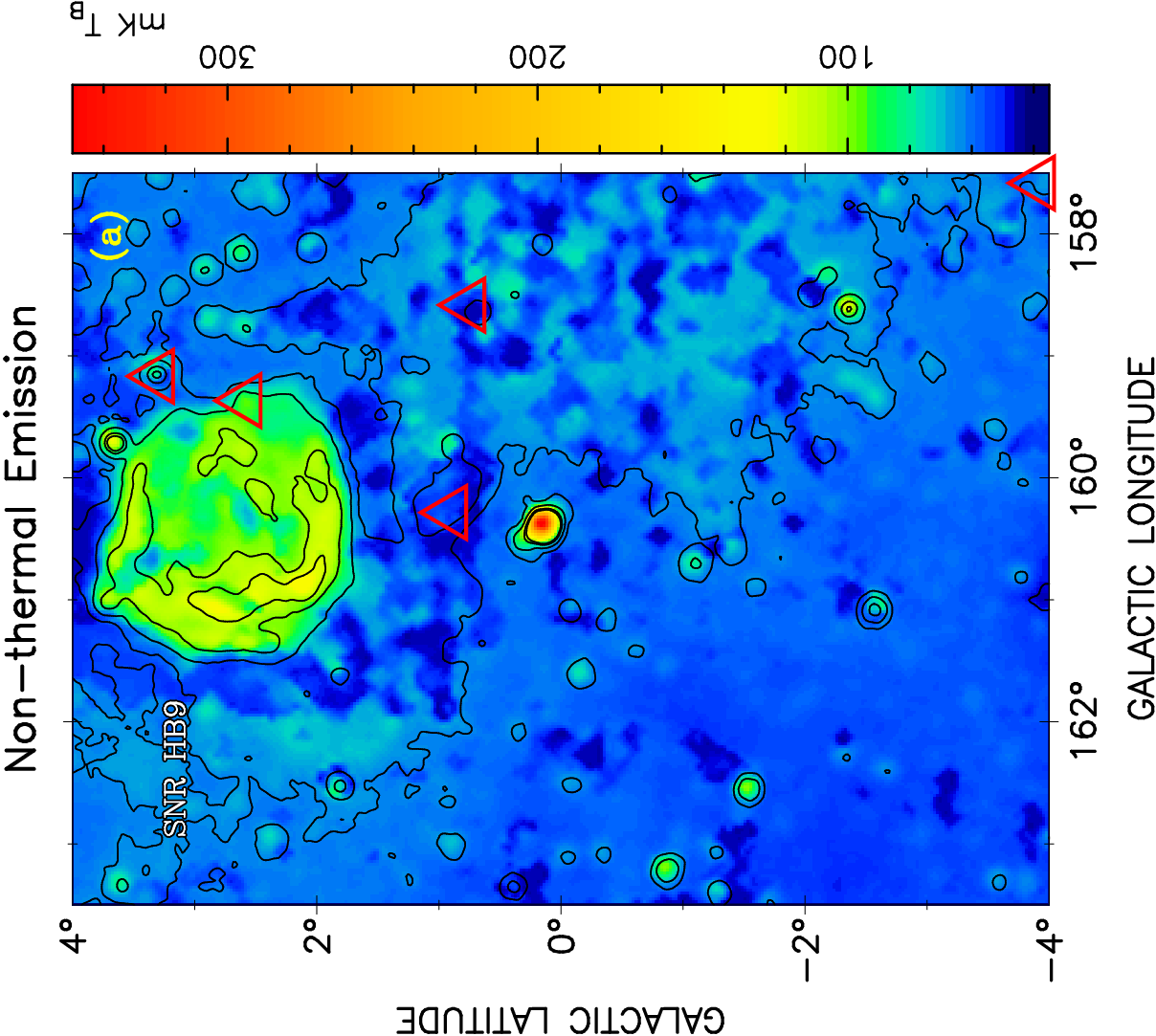}
\includegraphics[angle=-90,width=0.3\textwidth]{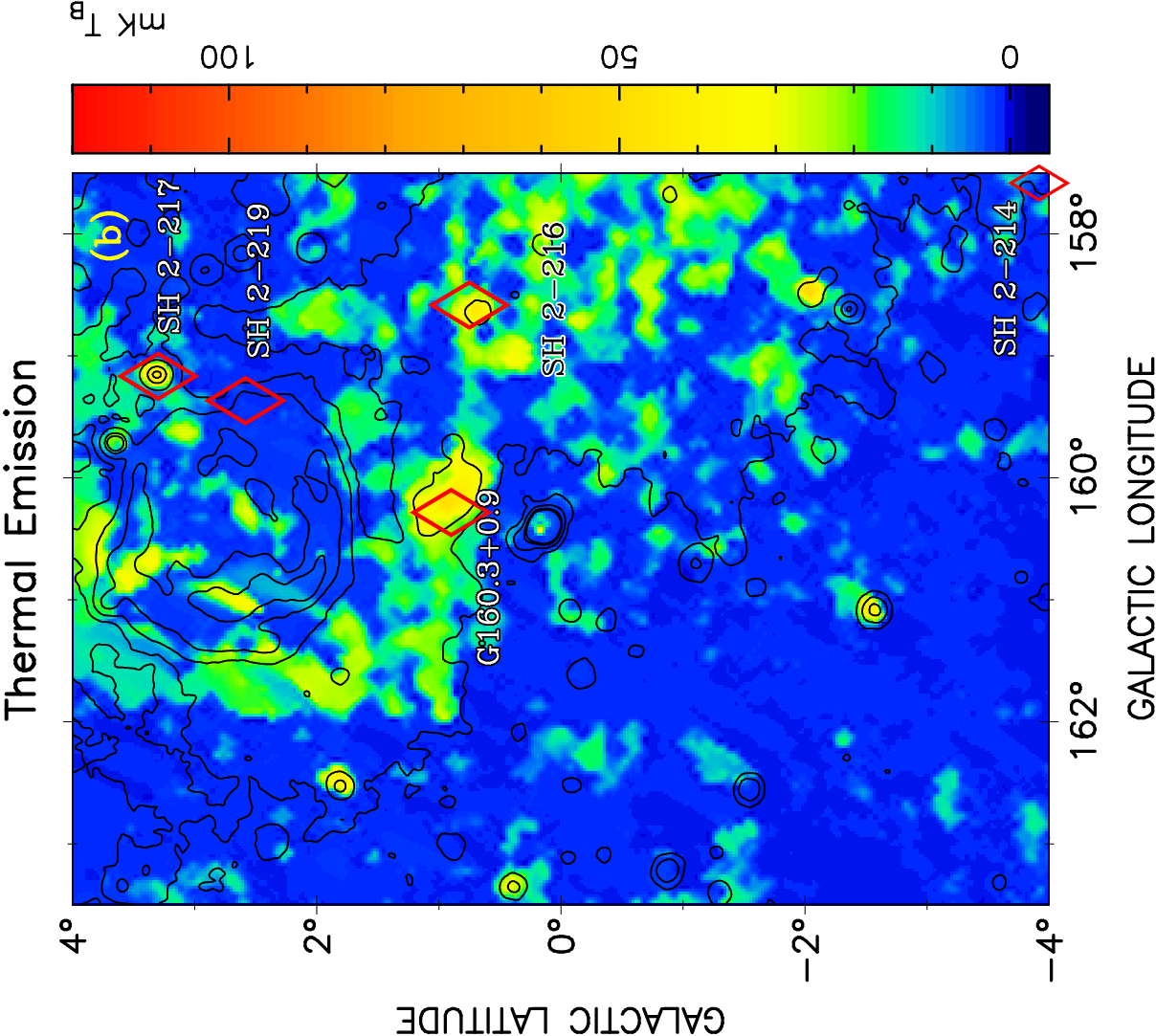}
\includegraphics[angle=-90,width=0.3\textwidth]{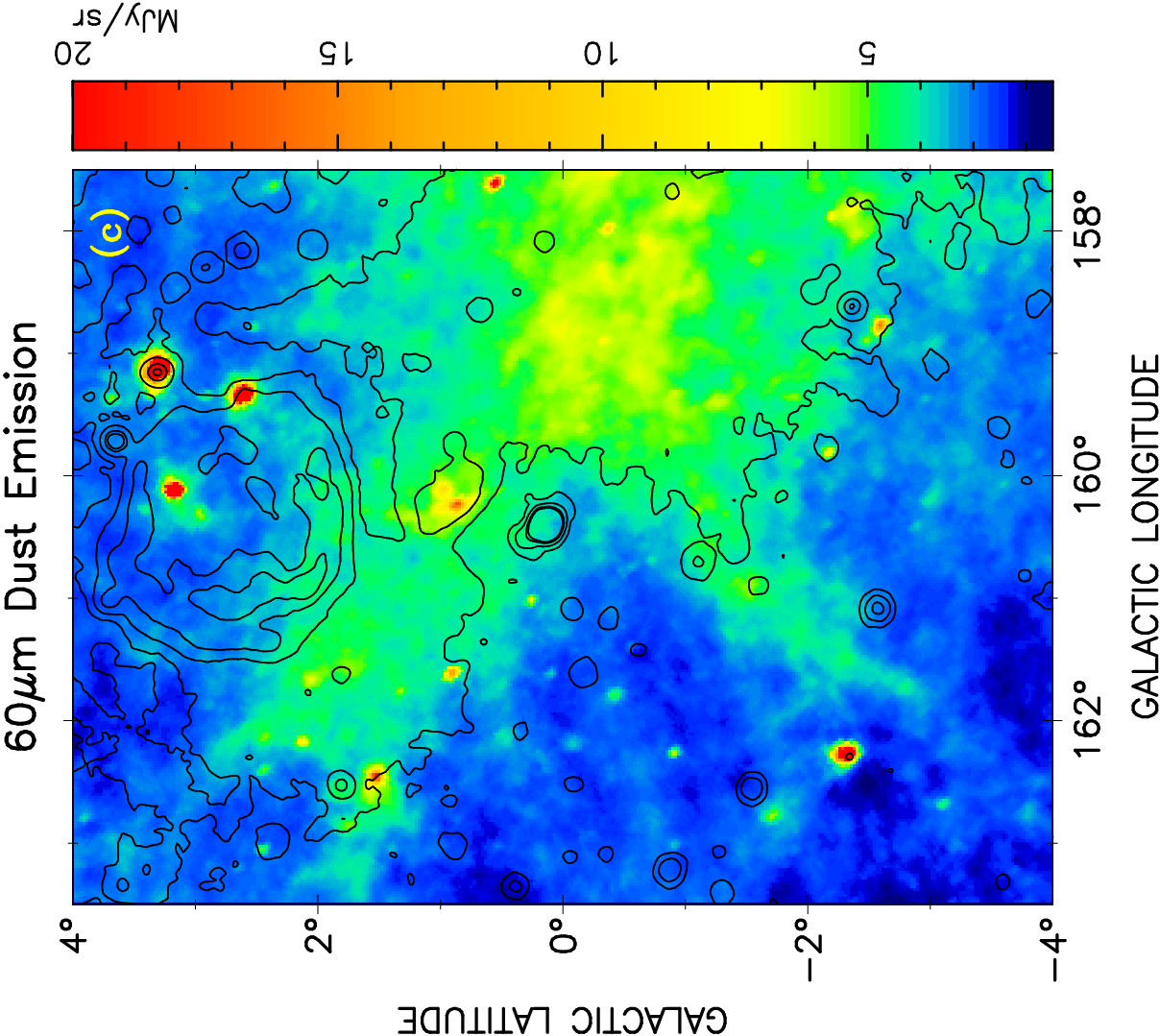}\\[5mm]
\includegraphics[angle=-90,width=0.24\textwidth]{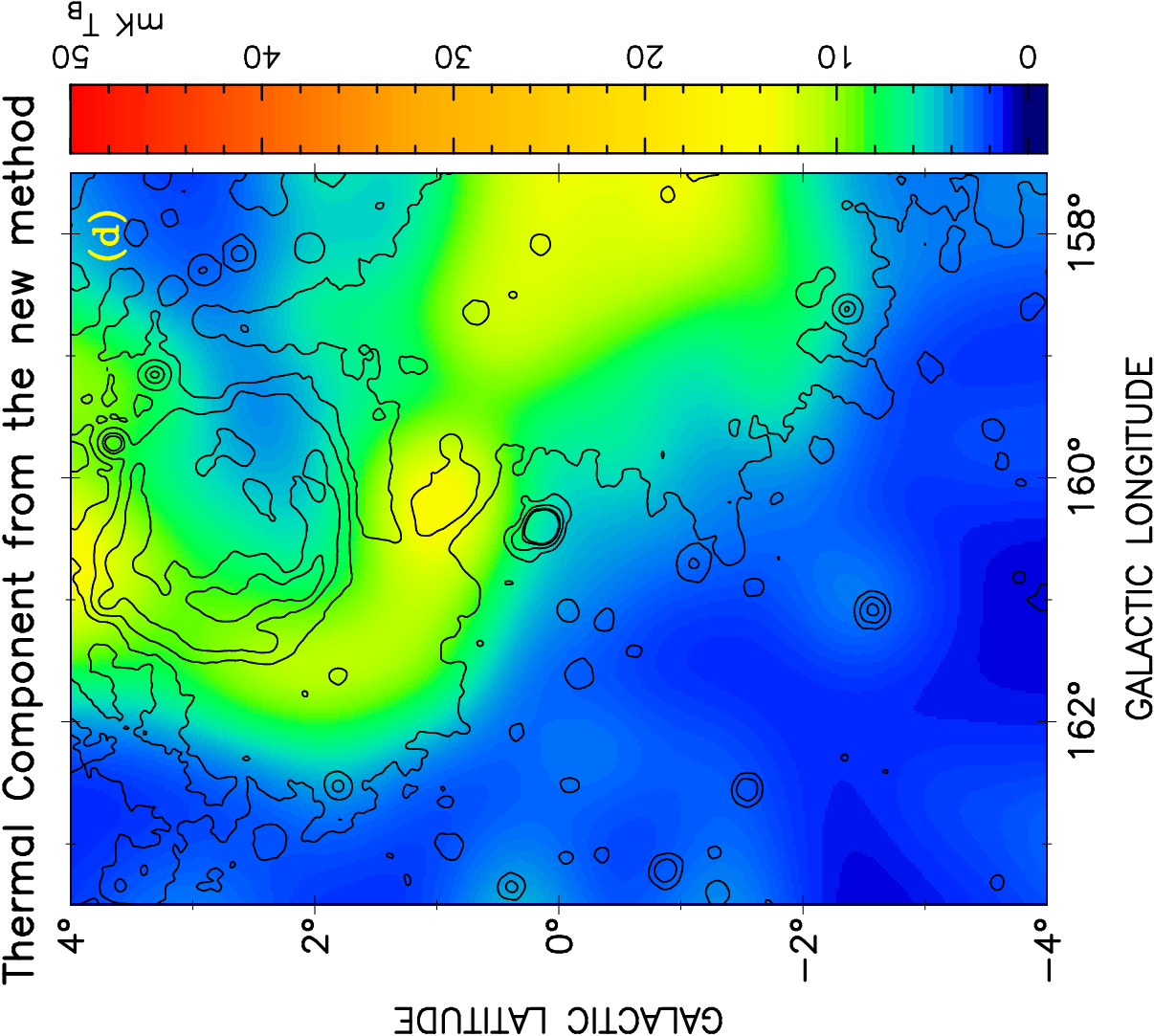}
\includegraphics[angle=-90,width=0.24\textwidth]{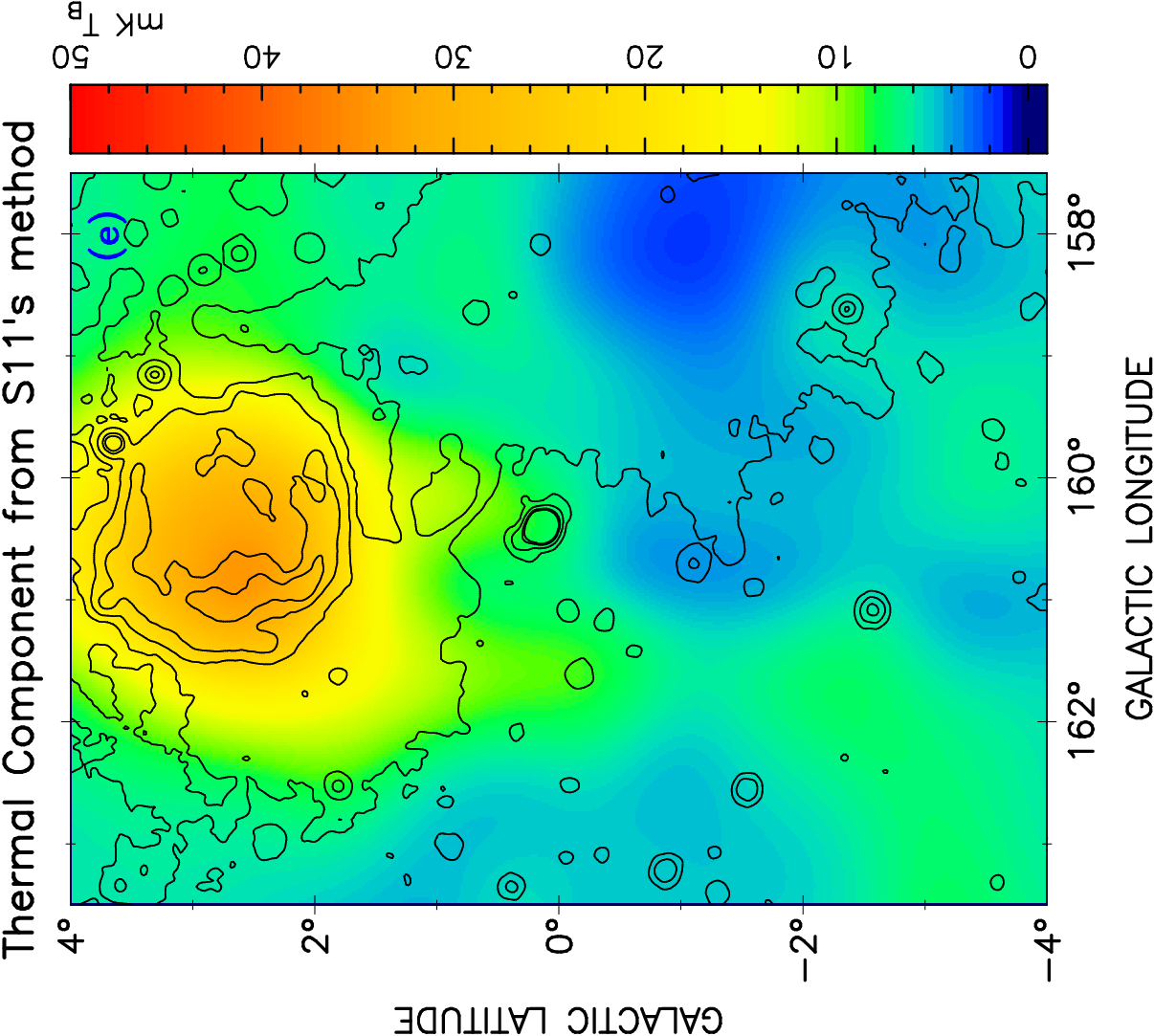}
\includegraphics[angle=-90,width=0.24\textwidth]{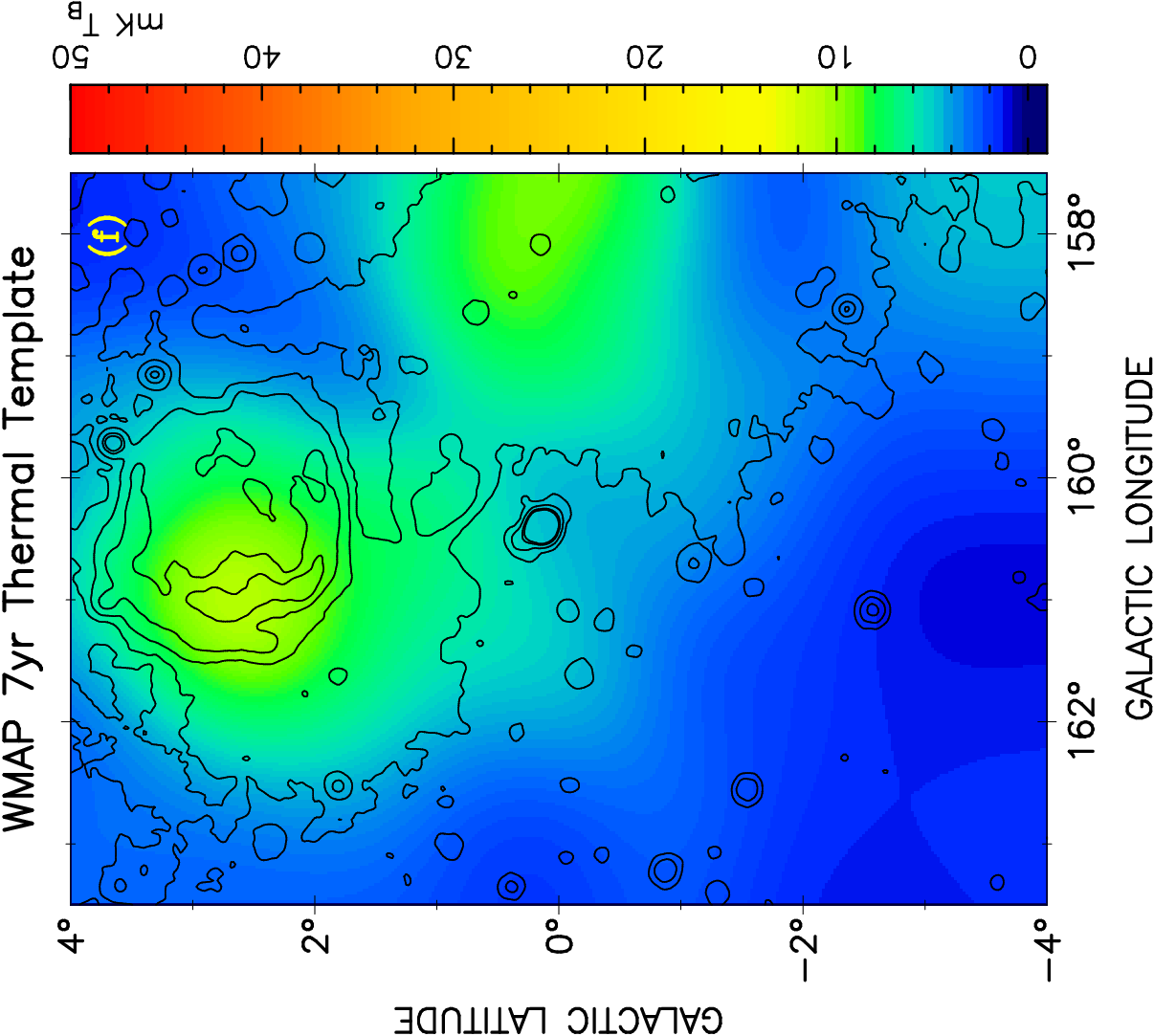}
\includegraphics[angle=-90,width=0.24\textwidth]{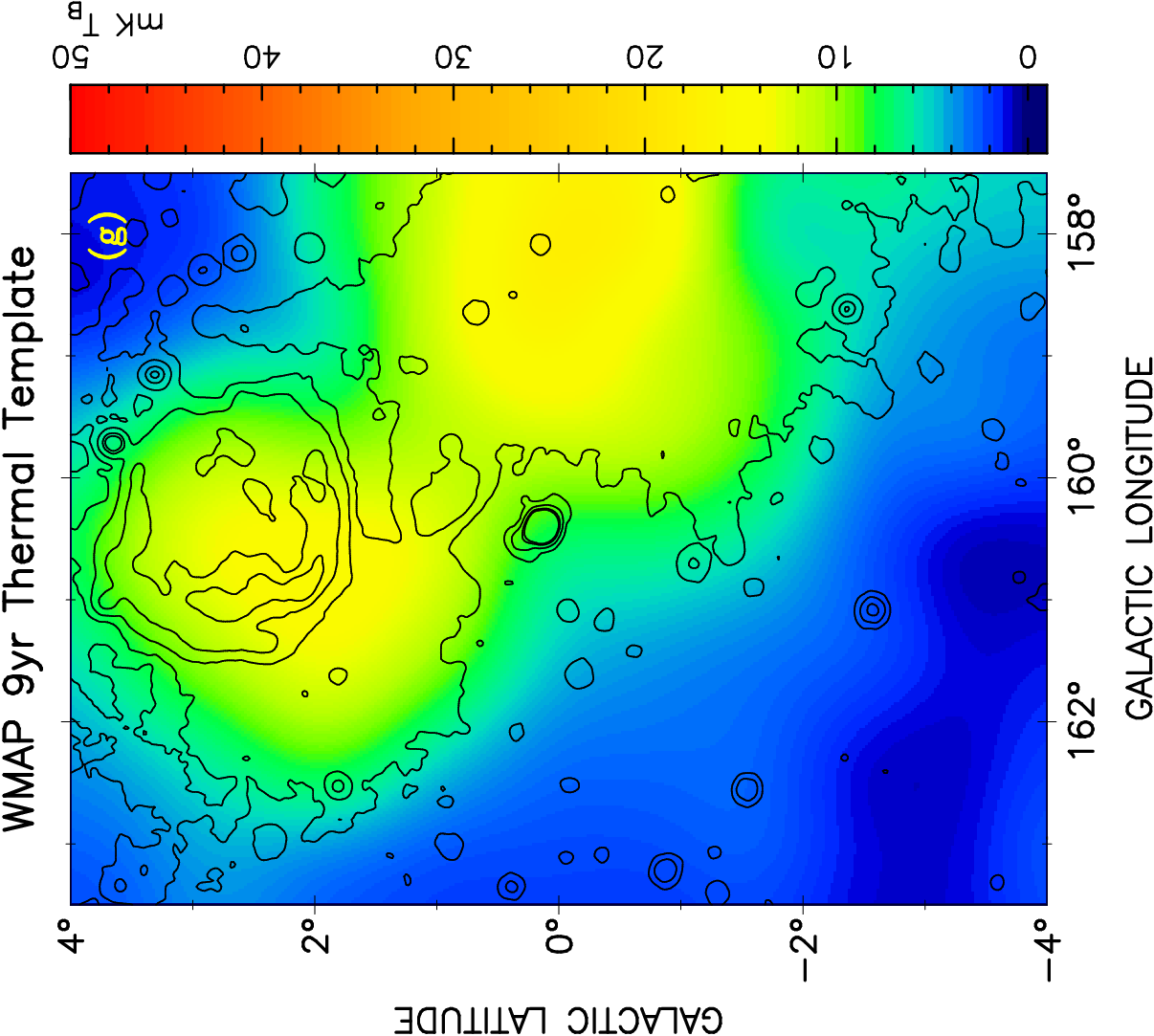}
\caption{Maps for the test region. {\it Upper panels:} high-resolution
  (9$\farcm$5) images of decomposed non-thermal ({\it panel a}) and
  thermal ({\it panel b}) emission at $\lambda$6\ cm overlaid by
  contours of the $\lambda$6\ cm total intensity (not decomposed). The
  contour lines run in 50, 65, 80, 105 and 120~mK T$_{b}$. The SNR HB9
  is labelled with name, while the known \ion{H}{II} regions are
  labelled with diamonds and names. {\it panel c:} the IRIS 60$\mu$m
  image overlaid by the same contours as in {\it panel a and
    b}. Images in the {\it lower panels} are convolved to an angular
  resolution of 1$\degr$, and overlaid by the same contours as in the
  {\it upper panels}. Decomposed thermal emission component by using
  our method ({\it panel d}) is compared with that by \citet{Sun11a}'s
  method ({\it panel e}) and WMAP 7-year and 9-year free-free
  templates ({\it panel f and g}).}
\label{hb9}
\end{figure*}

\noindent This equation is the key to all of the following
derivations. We explain our algorithm for the component separation in
below. We first subtracted $T_{off}$ in the 408~MHz, Stockert
1420~MHz, Effelsberg 1408~MHz maps and the dust
template\footnote{http://lambda.gsfc.nasa.gov/product/map/dr4/mem\_maps\_get.cfm}
from the WMAP K-band map. Then they are convolved to a common angular
resolution of 1$\degr$. According to Eq.~6, the 408~MHz and the
Stockert 1420~MHz data can be used to calculate the thermal fraction
at 1420~MHz, while the Effelsberg 1408~MHz and the WMAP 22800~MHz data
can be used for the derivation of the thermal fraction at
1408~MHz. The two fraction values should be the same at nearly the
same frequency. The spectal index of observed Galactic emission were
calculated pixel by pixel as $\beta_{408-1420}$ =
$\frac{log({T_{408}}/{T_{1420}})} {log({408}/{1420})}$ and
$\beta_{1408-22800}$ = $\frac{log({T_{1408}}/
  {T_{22800}})}{log({1408}/{22800})}$ according to the Eq.~5. The
thermal brightness temperature spectral index is set to a fixed value
of $\beta_{ff} = -2.1$. The only unknown non-thermal synchrotron
emission spectral index $\beta_{syn}$ in Eq.~6 is being fitted in the
range of $-$2.1 to $-$3.0. If the thermal fractions
($f_{th_{1.4GHz}}$) calculated from the 408/1420 data pair and the
1408/22800 data pair have a difference less than 5\%, the
$\beta_{syn}$ is recorded. Such a group of $\beta_{syn}$ is finally
averaged and taken as the final solution for the pixel. If no
$\beta_{syn}$ fits for a pixel, we set the thermal fraction
$f_{th_{1.4GHz}}$ as the average value of the surrounding pixels and
then calculate $\beta_{syn}$. If both $\beta_{408-1420}$ and
$\beta_{1408-22800}$ are larger than $-$2.1, the emission is
overwhelmingly contributed by the thermal emission. The non-thermal
emission of the pixel is taken as that from the nearest pixels where
the calculation is possible.  After this procedure, we got the map of
thermal fraction, and the brightness temperature map of thermal
emission at 1.4~GHz at an angular resolution of 1$\degr$. Non-thermal
emission map is then calculated by subtracting the thermal emission
map from the 1408~MHz total intensity map. With the spectral index
$\beta_{th} = -2.1$ and the $\beta_{syn}$ pixel by pixel, the thermal
and non-thermal emission components can be separated from observations
at any frequencies.

The separated components obtained above have an angular resolution of
1$\degr$. By taking the data from the Urumqi $\lambda$6\ cm,
Effelsberg $\lambda$11\ cm and the Effelsberg $\lambda$21\ cm
(1408~MHz) survey, a decomposition at an angular resolution 9$\farcm$5
can be made. However, as described in Sect.~2, both of the
$\lambda$6\ cm and $\lambda$11\ cm total intensity data miss the
large-scale emission component. $T_{zero}$ must be corrected in prior
to the separation. S11 developed a method for the base-level
restoration for the Urumqi survey towards the inner Galactic plane
area. Their method requires structureless edges of the map in the
latitude direction. It is not applicable to the Cygnus X region, since
the complex is too extended, exceeding 4$\degr$ in $b$. We therefore
restored the missing large-scale component for the $\lambda$6\ cm data
according to the method introduced by \citet{Reich9021}. We first
extrapolated the 1$\degr$ angular resolution Effelsberg 1408~MHz
thermal and non-thermal emission component to $\lambda$6\ cm by using
$\beta_{th}$ and $\beta_{syn}$, pixel by pixel. The sum of the two
extrapolated components was set as a fully calibrated template which
contains the emission from all scales. Then we convolved the
originally observed $\lambda$6\ cm data to the same angular resolution
of 1$\degr$ and compare with the template. The difference is
calculated and added back to the observed $\lambda$6\ cm data as the
missing large-scale component. The same procedure is done for the
Effelsberg $\lambda$11\ cm data. Then the Effelberg $\lambda$11\ cm
and $\lambda$21\ cm data were convolved to the angular resolution of
9$\farcm$5, the same as the Urumqi $\lambda$6\ cm data. Following the
algorithm we introduced above, component separation with a higher
angular resolution can be realized at the wavelength of
$\lambda$6\ cm.

\subsection{Results for a test region}

In order to verify the capability of our new method, a region of
$158.5\degr \leqslant \ell \leqslant 163.5\degr$, $|b| \leqslant
4\degr$, involving the well-known SNR HB9, some \ion{H}{II} regions
and diffuse emission was firstly tested qualitatively. We showed the
decomposition results in the {\it upper panel} of the
Fig.~\ref{hb9}. The maps show reasonable results for many sources. The
SNR HB9 and many extra-Galactic sources e.g. four extra-Galactic
sources located around $\ell \sim 158\fdg5, b \sim 2\fdg7$ in the west
of HB9, the sources at $\ell = 163\fdg35, b = 0\fdg35$, $\ell =
163\fdg20, b = -0\fdg90$, $\ell = 162\fdg55, b = -1\fdg55$, and $\ell
= 160\fdg70, b = -1\fdg15$ clearly show up in the separated
non-thermal component image (Fig.~\ref{hb9}, {\it panel a}) as
expected. Free-free emission of the large \ion{H}{II} region SH 2-216
\citep[1$\fdg$6 in diameter][]{Tweedy95}, and the \ion{H}{II} region
G160.3+0.9 \citep{Blitz82, Kuchar97} are successfully separated into
the thermal component map (Fig.~\ref{hb9}, {\it panel b}).  Thermal
emission associate with the \ion{H}{II} region SH 2-214 ($\ell =
157\fdg60, b = -3\fdg91$) is not seen, since no clear radio
counterpart of this object was detected at $\lambda$6\ cm.

\begin{figure*}
\centering
\includegraphics[angle=-90, width=0.8\textwidth]{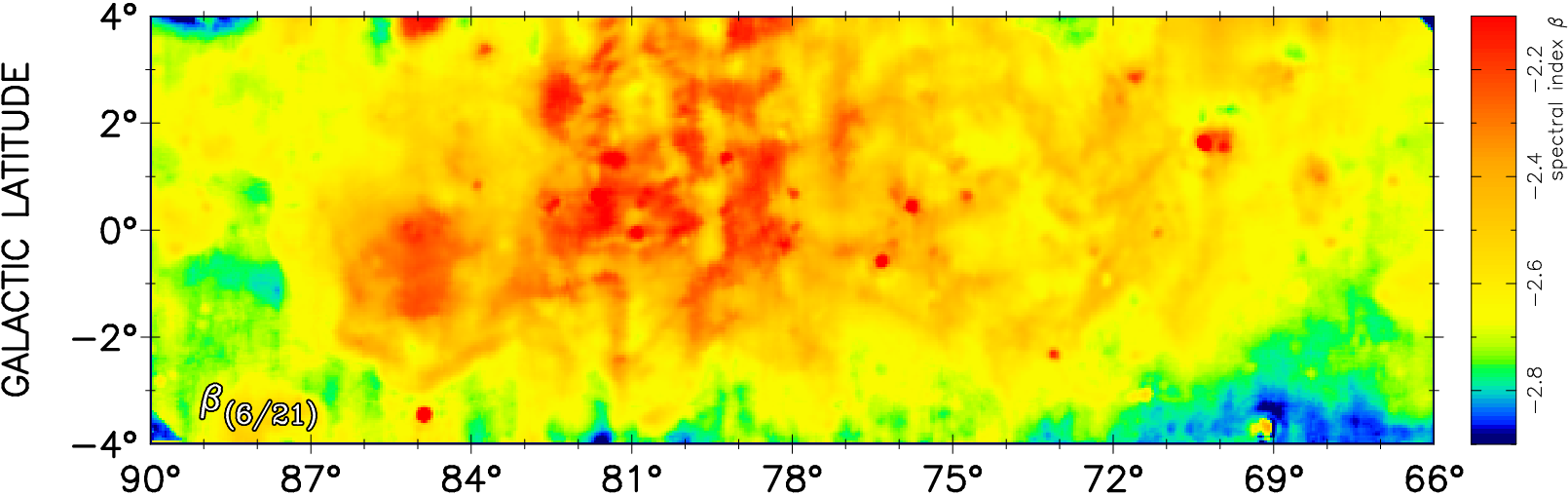} \\
\includegraphics[angle=-90, width=0.8\textwidth]{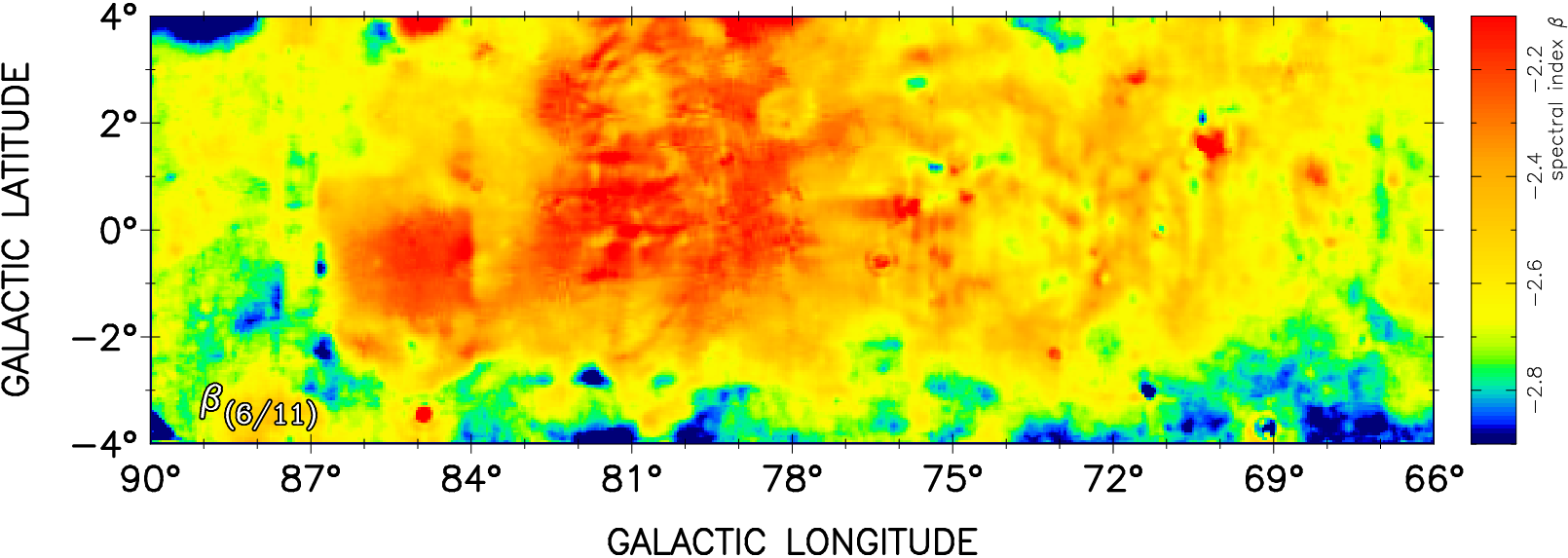} \\
\caption{Spectral index maps of the Galactic emission derived from
  survey maps at $\lambda$6 cm and $\lambda$21 cm bands ({\it upper
    panel}) and at $\lambda$6\ cm and $\lambda$11\ cm bands ({\it
    lower panel}).}
\label{spectra}
\end{figure*}

\begin{figure*}[!th]
\centering
\includegraphics[angle=-90,scale=0.55]{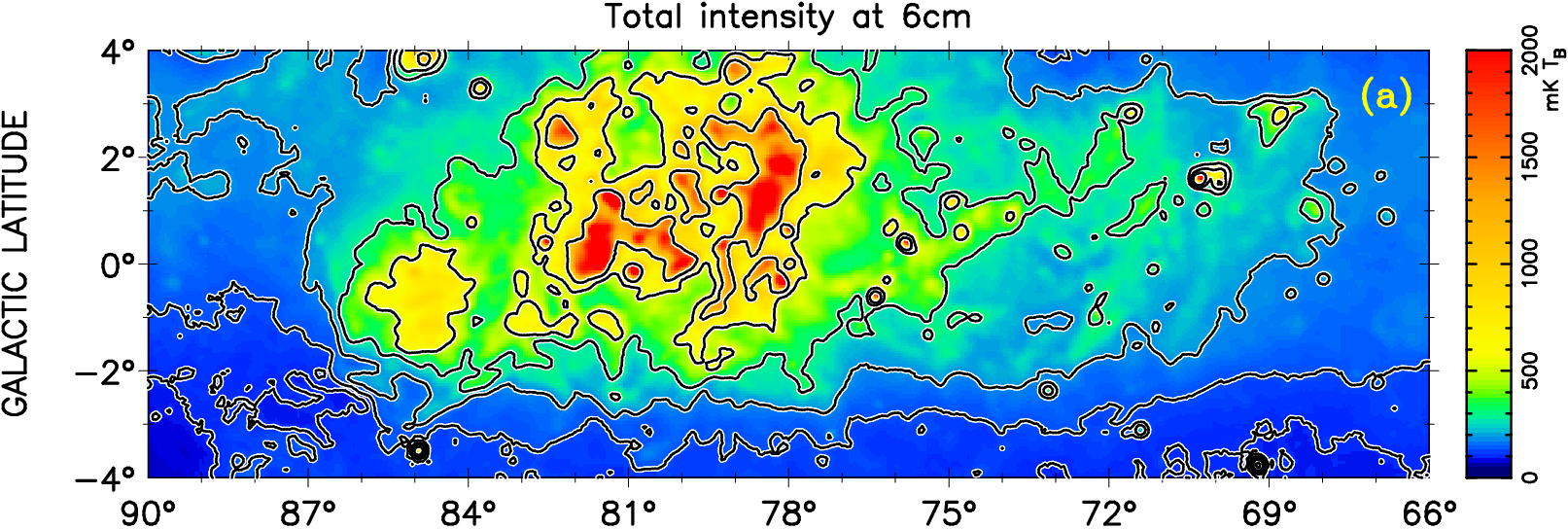} \\[4mm]
\includegraphics[angle=-90,scale=0.55]{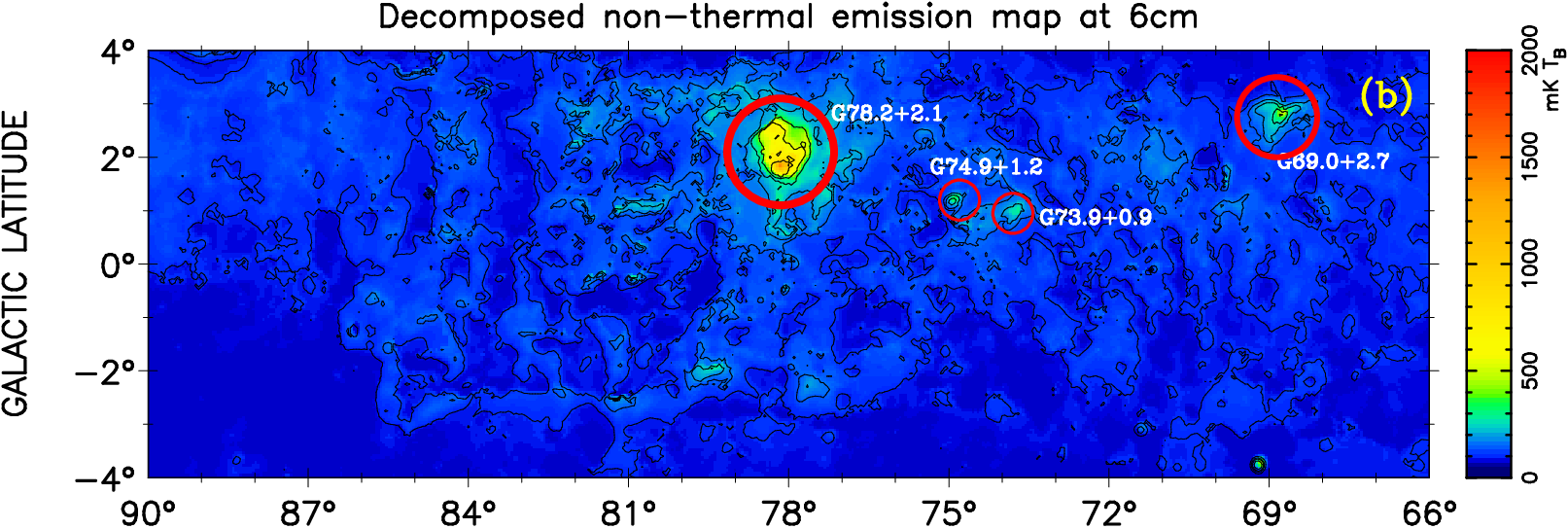} \\[4mm]
\includegraphics[angle=-90,scale=0.55]{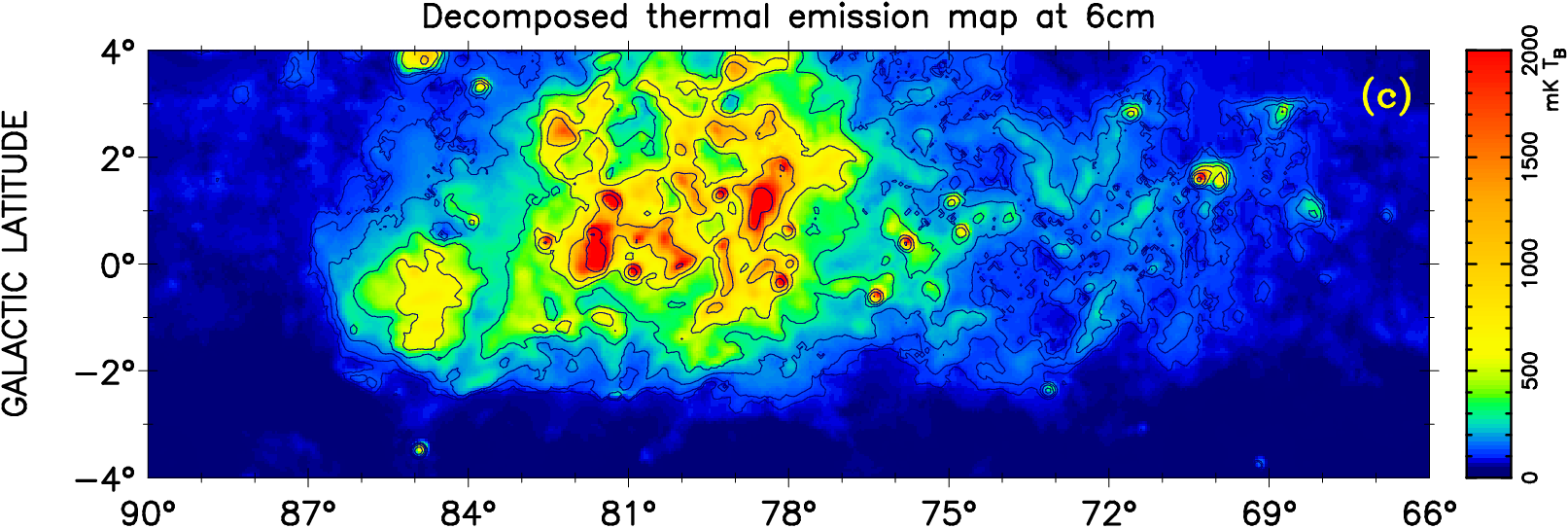} \\[4mm]
\includegraphics[angle=-90,scale=0.55]{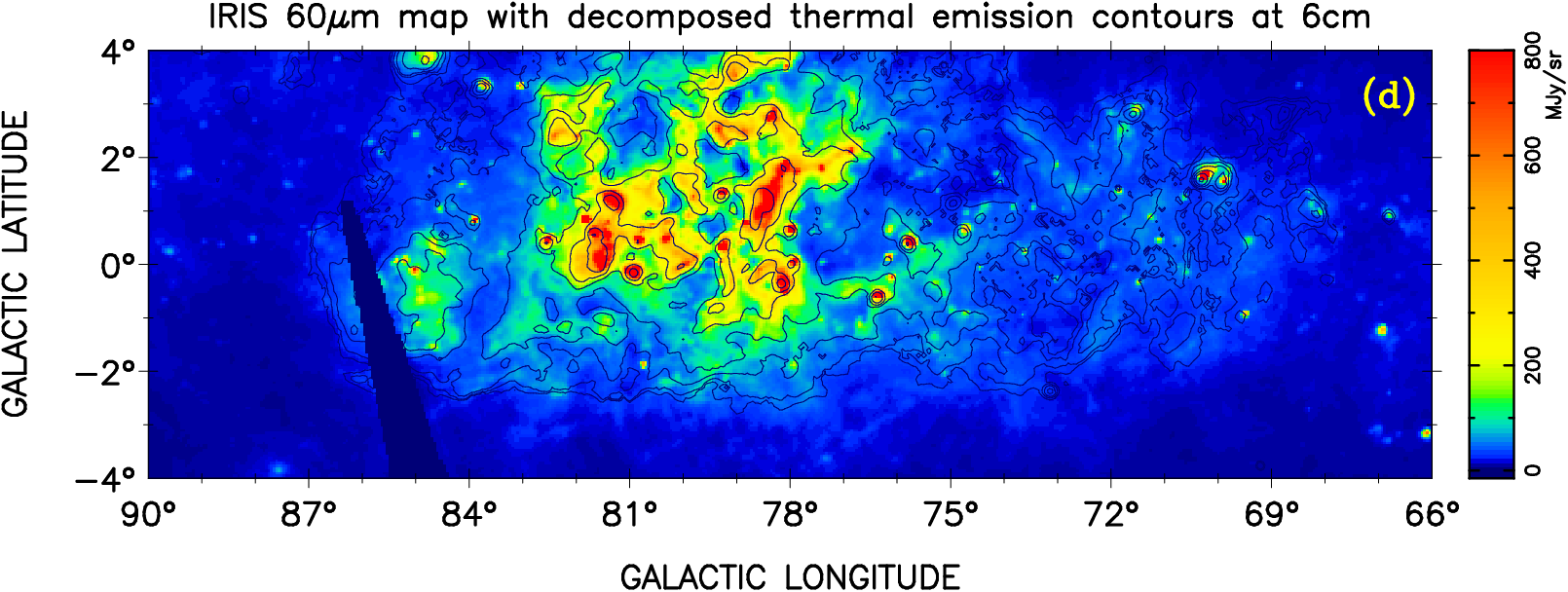} \\[4mm]
\caption{The $\lambda$6\ cm radio maps of total intensity ({\it panel
    a}), decomposed non-thermal ({\it panel b}) and thermal ({\it
    panel c}) emission in the Cygnus X region at an angular resolution
  of 9$\farcm$5. The IRIS 60$\mu$m image ({\it panel d}) of the Cygnus
  X region is shown for comparison with the decomposed thermal
  map. Contours in {\it panel a and c} are running in steps of
  $70+2^{n-1}\times30$~mK T$_{b}$ (n= 0, 1, 2...). The contours in
  {\it panel b} for non-thermal emission run in 100, 150, 200, 300,
  500 and 800~mK T$_{b}$. Contours in {\it panel d} are the same as
  those in {\it panel c}. The visible SNRs are indiacted in {\it panel
    b}.}
\label{CygnusX}
\end{figure*}

We further cross check the output of the new method with that from
S11's method, and those from the free-free emission template of the
7-year \citep{Gold11} and 9-year WMAP data \citep{Bennett12}
extrapolated from $\lambda$1.3\ cm to $\lambda$6\ cm with the
brightness temperature spectral index of $\beta = -2.1$, at the same
angular resolution of 1$\degr$. It is important and necessary to do
these comparisons, since our method and that of S11 both rely on the
radio continuum data and the determination of the non-thermal spectral
indices, while the WMAP template is independent, using the H$\alpha$
data to estimate the contribution from the thermal free-free
emission. The comparison is shown in the lower panels of the
Fig.~\ref{hb9} and we found our result is better. Generally, the
thermal emission component map produced by our new method show similar
structures as shown in the WMAP templates. However, the images from
the WMAP templates and the S11 method unexpectedly show a large amount
of thermal emission within the SNR HB9, which is supposed to be a
purely non-thermal source. S11 explained this leakage as a result from
a fixed non-thermal synchrotron spectral index. For the large thermal
\ion{H}{II} region SH 2-216, we found structural resemblance between
radio emission and the IRIS 60$\mu$m infrared dust emission
\citep{Miville05}, which confirms the thermal property of the large
area. The thermal emission of SH 2-216 is well separated out by our
method, and also shown in the two WMAP templates, but very little
thermal emission can be separated out by S11's method. We noticed that
the thermal emission of SH 2-216 figured out by our method is less
than that in the WMAP 9-year template, but more than that in the WMAP
7-year template. We do not understand the big difference between the
two templates.

Another indication of validation of the new method comes from an
identical structure to the southeast of the SNR HB9. Radio emission in
this region can be well separated in the thermal component map and
seen in the images shown in Fig.~\ref{hb9} by all of the four
methods. Strong infrared emission in this region is detected in the
IRIS 60$\mu$m image, implying its thermal nature.

``Leakage'', however, is still seen in a few sources, perhaps due to
the fluctuations in the data pixels from the observations. For
example, the thermal \ion{H}{II} region SH 2-217 ($\ell = 159\fdg16, b
= 3\fdg30$) is partially seen in the non-thermal emission image.

\section{Application of the new method to the Cygnus X region}

After proving the applicability of the new method, we apply it to
decompose the thermal and non-thermal emission components in the
Cygnus X region. As noticed above that the WMAP 7-year and 9-year data
show discrepancies in the test region. We involved the two versions of
the WMAP data in our calculation alternatively. The 7-year data gives
more reasonable result and is finally used. We show in
Fig.~\ref{spectra} the spectral index $\beta$ maps of the observed
Galactic emission in the Cygnus X region derived between the Urumqi
$\lambda$6\ cm and Effelsberg $\lambda$21\ cm (1408~MHz) data, and
between the Urumqi $\lambda$6\ cm and Effelsberg $\lambda$11\ cm data,
and present in Fig.~\ref{CygnusX} the separation result of the thermal
and non-thermal components. The spectral index map derived between
$\lambda$6\ cm and $\lambda$21\ cm is much smoother than that from
$\lambda$6\ cm and $\lambda$11\ cm. We suspect that it is partially
due to the small frequency separation and partially due to scanning
effect in the $\lambda$11\ cm survey, shown as horizontal stripes
around $\ell \sim 81\degr$ in the separation results as shown in
Fig.~\ref{spectra}. The first glance of the overall structure
decomposed by the new method confirms the conclusions of
\citet{Wendker91} that nearly all the prominent radio radiation from
the Cygnus X region comes from the thermal free-free emission. Many
elongated thermal ridges are identical to what they have
discovered. All known \ion{H}{II} regions listed by
\citet{Sharpless59} and \citet{Paladini03} are found to have thermal
emission in the separated image, which can be verified by the IRIS
60$\mu$m image (see Fig.~\ref{CygnusX}, {\it panel d}).

\subsection{Diffuse emission}

\begin{figure}
\centering
\includegraphics[angle=-90,scale=0.4]{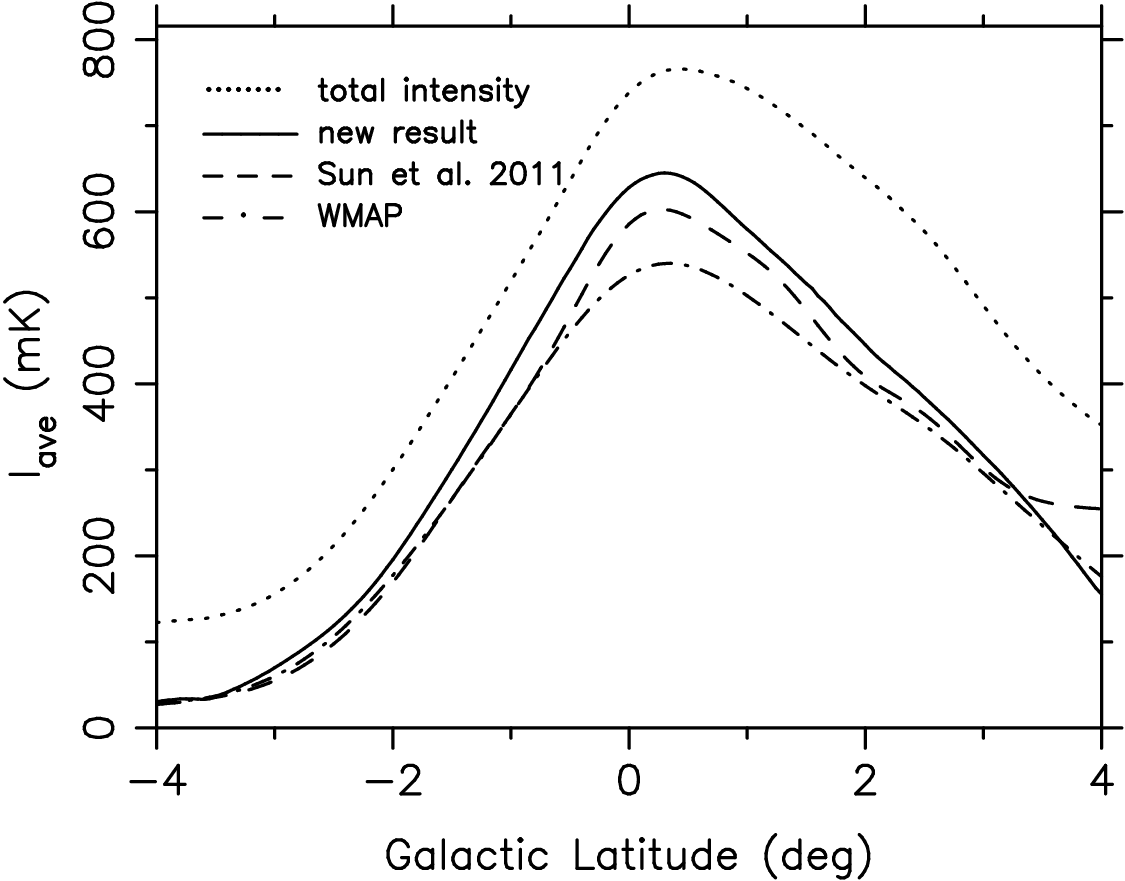}
\includegraphics[angle=-90,scale=0.4]{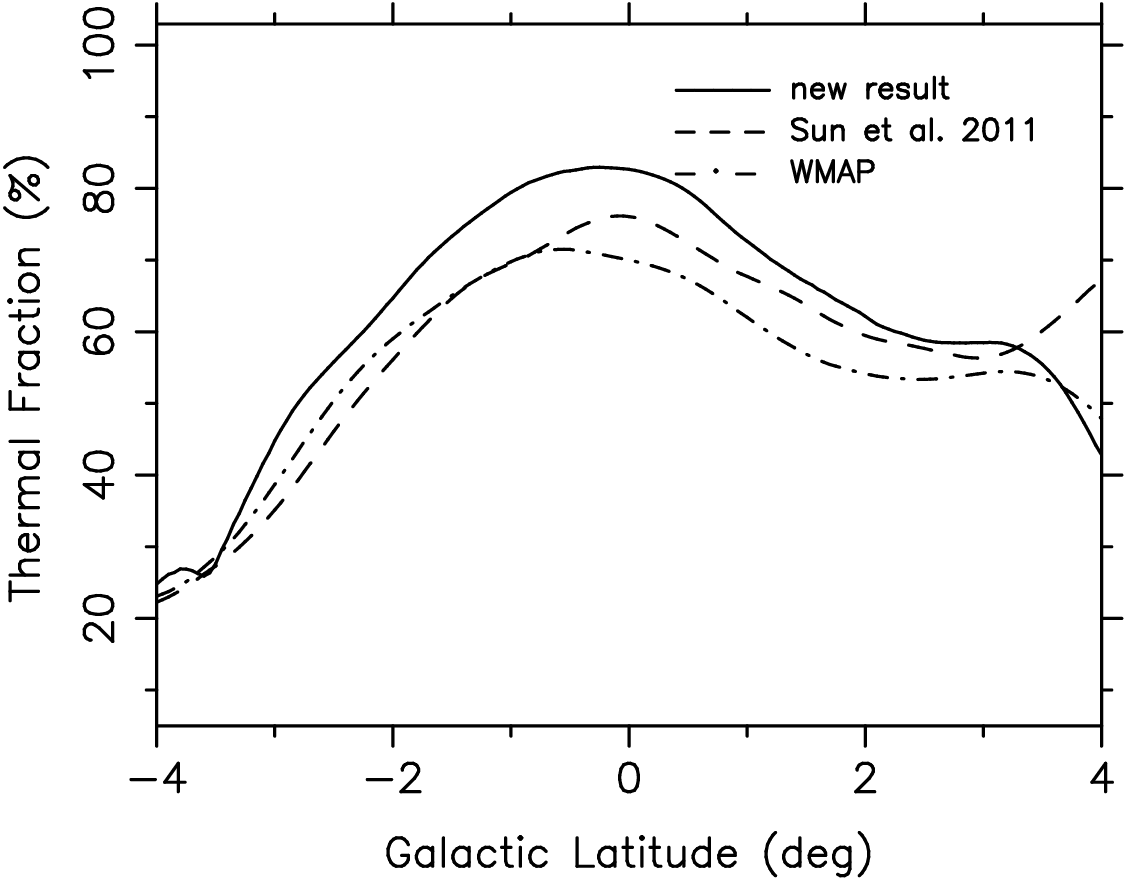}
\caption{Galactic latitude profiles for the averaged intensities ({\it
    upper panel}) and averaged thermal fractions ({\it lower panel})
  at $\lambda$6\ cm in the Cygnus X region ($77\degr \leqslant \ell
  \leqslant 87\degr, |b| \leqslant 4\degr$). The dotted line is the
  $\lambda$6\ cm total intensity. The thermal component derived from
  our method, S11's method and WMAP template are presented by solid
  line, dashed line and dash-dot line, respectively.}
\label{cygfra}
\end{figure}

Non-thermal radiation in the Cygnus X region is weak and diffuse (see
Fig.~\ref{CygnusX}, {\it panel b}). No significant enhancement in
synchrotron radiation across the longitude range of 66$\degr$ to
90$\degr$ is noticed. This does not conflict with the point of view
that the Cygnus X region is a complex that the local arm is seen
end-on \citep{Wendker91}. Either because the local arm has a certain
width and our angle of view might not exceed this scale and/or the
local arm is relatively fainter than the major arms, that the
non-thermal radiation we see in this region is predominately
contributed by the farther Perseus arm.  The r.m.s of the non-thermal
component image is about 20~mK T$_{b}$ at $\lambda$6\ cm. Although the
image appears to be mottled, which we believe was introduced by the
uncertainties in the data, the intensity in large scales is uniform in
general, around 100~mK T$_{b}$ at $\lambda$6\ cm. This quantity, shown
as the gap between the total power and thermal emission profiles in
Fig.~\ref{cygfra} is always standing there even by the three different
methods. With the non-thermal emission components separated at
$\lambda$21\ cm, $\lambda$11\ cm and $\lambda$6\ cm, discarding the
regions of the known SNRs, we estimate the average synchrotron
spectral index of the Cygnus X region as being $\beta_{syn} =
-2.8\pm0.1$.

The distribution of thermal radiation concentrates in two areas (see
Fig.~\ref{CygnusX}, {\it panel c}). One is the main part of the Cygnus
X region, in the centre of the map, about 9$\degr$ span in the
Galactic longitude direction. The majority of known \ion{H}{II}
regions can be found in this region. The other area with enhanced
thermal emission is W80 ($\ell \sim 85\degr, b \sim -0\fdg9$), whose
optical counterpart contains the well-known North America and the
Pelican Nebulae. The thermal radiation in both of the areas account
for $\sim$ 75\% of the total intensity of the $\lambda$6\ cm continuum
emission.

For the large-scale thermal emission, we statistically compared the
one-dimensional averaged results derived by our method, S11's method
and the WMAP template at $\lambda$6\ cm at an angular resolution of
1$\degr$. The intensity of each pixel is averaged in the longitude
direction and the profiles are shown in Fig.~\ref{cygfra}. The three
profiles highly resemble each other and the decomposition failures in
the individual sources (like the case of SNR HB9) do not affect the
statistics on very large scales, as noted by S11.

\subsection{Uncertainty of results}

A qualitative comparison in Fig.~\ref{hb9} shows that the new
separation method performances better in details than the other
methods. In a statistical sense, the new method gives consistent
results with that by the WMAP template and the S11's method (see
Fig.~\ref{cygfra}). Here we estimate of the uncertainties of the
separation result of the Cygnus X region. From the errors of the low
resolution surveys, an uncertainty of 5\% was esitamted for the
$\lambda$6\ cm and $\lambda$11\ cm data after the restoration of the
missing large-scale emission. Based on the standard error propagation,
the uncertainty of the separated thermal and non-thermal components
derived by the new method with the Effelsberg $\lambda$21\ cm,
$\lambda$11\ cm and the Urumqi $\lambda$6\ cm data as we showed in the
Fig.~\ref{CygnusX} is about 17\%. As noted by \citet{Wendker91}, the
accuracy in the low emission area, usually the data points away from
the Galactic plane near edges of the image, tends to be worse than
that for the area of strong emission with high S/N ratios. We tested
the pixels near the boundary of the separated map, and found the
uncertainties are indeed higher, but at most 30\%.

\subsection{Notes on discrete sources}

According to \citet{Paladini03}, 145 individually known \ion{H}{II}
regions are located in the Cygnus X region. We summed their flux
density according to the Paladini catalogue and extrapolate it to
$\lambda$6\ cm band by the spectral index of $\beta = -2.1$. We found
that the individual \ion{H}{II} regions only account for 4\% of the
total thermal emission at $\lambda$6\ cm. This might be the lower
limit, since \ion{H}{II} regions are difficult to identify in such a
complex area. However, undoubtedly, diffuse thermal emission
overwhelmingly dominates the whole thermal emission budget.

There are 12 known supernova remnants in the catalog of
\citet{Green09} in the region ranging of $66\degr \leqslant \ell
\leqslant 90\degr$, $|b| \leqslant 4\degr$. Two large SNRs, G78.2+2.1
(DR4) and G69.0+2.7 (CTB~80) have already been studied in
\citet{Gao11x} with the Urumqi $\lambda$6\ cm survey data, while five
small SNRs, i.e. G73.9+0.9, G74.9+1.2, G76.9+1.0, G84.2$-$0.8 and
G85.9$-$0.6 were investigated in \citet{Sun11b}.  From the non-thermal
emission component shown in Fig.~\ref{CygnusX}, both SNR G78.2+2.1 and
CTB~80 are clearly separated out by the new method. In the original
map, the SNR G78.2+2.1 is heavily confused by the strong thermal
emission and the surrounding structures. \citet{Gao11x} used the
``background filtering'' technique \citep{Sofue79} and subtracted a
fitted twisted hyper plane to eliminate the un-related non-thermal
background and the ambient thermal emission \citep[see Fig.~1
  of][]{Gao11x}. The image of Fig.~\ref{CygnusX} (b) from the new
method show less contamination of thermal emission by successfully
getting rid of the confusion from the \ion{H}{II} region G78.3+2.8 in
the north, IC~1318b in the south and southeast, and perhaps also the
contribution of the un-related small \ion{H}{II} region, $\gamma$
Cygni nebula, which coincides in the southern shell of the SNR
G78.2+2.1. The $\lambda$6\ cm integrated flux density for this SNR is
now found to be $\sim$ 140~Jy, consistent with the result found by
\citet{Gao11x}. Faint non-thermal radiation is seen in the south,
north-west and north-east of the SNR G78.2$+$2.1. We suppose that
these might result from ``leakages''.

Besides the two large SNRs, we can see the SNRs G73.9+0.9, G74.9+1.2,
G84.2$-$0.8 and G85.9$-$0.6 from the separated map. The later two are
much fainter than their measured intensity \citep{Sun11b}. This is not
the ``leakage'' problem but due to the data we used. The spectral
indices $\beta$ of the two SNRs derived between $\lambda$6\ cm and
$\lambda$11\ cm are both around $\sim$ $-$2.2, too flat for shell-type
SNRs. We cannot separate out the SNR G76.9+1.0, since no solution can
be matched between the data pair of 1408~MHz/4800~MHz and
2700~MHz/4800~MHz.  The remaining five SNRs G67.7+1.8, G68.6$-$1.2,
G69.7+1.0, G85.4+0.7 and G83.0$-$0.3 were not included in any previous
Urumqi $\lambda$6\ cm studies. All of them are faint and less than
30$\arcmin$ in diameter, structureless as seen with our 9.5$\arcmin$
beam. The first three were deeply embedded in the radio complex of
W80. We do see some signatures of these objects in the decomposed
non-thermal emission image, however, they are fragmented. ``Leakages''
to the thermal emission with various amounts are detected for all of
them, thus no good measurement can be made.

Apart from the known SNRs, we carefully checked the spectral index
distributions of observed Galactic and the suspicious structures in
the non-thermal emission image for possible new SNRs, but failed to
found any new features. We estimate the surface brightness limit of
the decomposed non-thermal emission map by using the equation $\rm
\Sigma_{1GHz} = 1.505 \times 10^{-19} S_{1GHz} / \theta^{2}$, here
S$_{\rm 1GHz}$ is obtained by extrapolating S$_{6cm}$ = 1~Jy (160~mK),
the 3$\sigma$ level above the non-thermal background emission (100~mK)
to 1~GHz, using the typical spectral index of $\alpha = -0.5$ for
shell-type SNRs, $\theta$ is the beam size of 9$\farcm$5, the smallest
scale of sources that can be seen in the data. We finally conclude
that no new large-extent SNRs brighter than $\rm \Sigma_{1GHz} = 3.7
\times 10^{-21}\ W\ m^{-2}\ Hz^{-1}\ sr^{-1}$ can be found in our
separated non-thermal image of the Cygnus X region.

\section{Summary}

We developed a new method which can be used to separate the thermal
free-free and the non-thermal synchrotron emission components by using
the centimetre radio continuum survey data. The new method is applied
to the Cygnus X complex region. We found that the thermal free-free
emission comprises 75\% of the total continuum radiation in the Cygnus
X region at $\lambda$6\ cm. We compared the large-scale thermal
structures decomposed by the new method with that derived from the
WMAP data, and found that they are consistent with each other. An
uniform non-thermal background in the Cygnus X region is found to be
100$\pm$20~mK at $\lambda$6\ cm. The separation by using the
Effelsberg $\lambda$21\ cm, $\lambda$11\ cm and the Urumqi
$\lambda$6\ cm maps enables us to search for new large and faint SNRs
in the Cygnus X region at an angular resolution of
9$\farcm$5. However, no new large-extent SNRs brighter than $\rm
\Sigma_{1GHz} = 3.7 \times 10^{-21}\ W\ m^{-2}\ Hz^{-1}\ sr^{-1}$ are
discovered.

\begin{acknowledgements}

The authors would like to thank the anonymous referee for constructive
suggestions, and Dr. Wolfgang Reich and Dr. Xiaohui Sun for helpful
discussions. XWF would like to thank Mr. Tao Hong in helping with the
production of plots and carefully reading the manuscript.
  The authors are supported by the National Natural Science foundation
  of China (10773016, 11303035 and 11261140641) and XYG is
  additionally supported by the Young Researcher Grant of National
  Astronomical Observatories, Chinese Academy of Sciences.

\end{acknowledgements}

\bibliographystyle{aa}
\bibliography{bbfile}

\begin{thebibliography}{41}
\expandafter\ifx\csname natexlab\endcsname\relax\def\natexlab#1{#1}\fi

\bibitem[{{Alves} {et~al.}(2012){Alves}, {Davies}, {Dickinson}, {Calabretta},
  {Davis}, \& {Staveley-Smith}}]{Alves12}
{Alves}, M.~I.~R., {Davies}, R.~D., {Dickinson}, C., {et~al.} 2012, \mnras,
  422, 2429

\bibitem[{{Bennett} {et~al.}(2003){Bennett}, {Halpern}, {Hinshaw}, {Jarosik},
  {Kogut}, {Limon}, {Meyer}, {Page}, {Spergel}, {Tucker}, {Wollack}, {Wright},
  {Barnes}, {Greason}, {Hill}, {Komatsu}, {Nolta}, {Odegard}, {Peiris},
  {Verde}, \& {Weiland}}]{Bennett03}
{Bennett}, C.~L., {Halpern}, M., {Hinshaw}, G., {et~al.} 2003, \apjs, 148, 1

\bibitem[{{Bennett} {et~al.}(2012){Bennett}, {Larson}, {Weiland}, {Jarosik},
  {Hinshaw}, {Odegard}, {Smith}, {Hill}, {Gold}, {Halpern}, {Komatsu}, {Nolta},
  {Page}, {Spergel}, {Wollack}, {Dunkley}, {Kogut}, {Limon}, {Meyer}, {Tucker},
  \& {Wright}}]{Bennett12}
{Bennett}, C.~L., {Larson}, D., {Weiland}, J.~L., {et~al.} 2012, \apjs, accepted [arXiv:1212.5225]

\bibitem[{{Blitz} {et~al.}(1982){Blitz}, {Fich}, \& {Stark}}]{Blitz82}
{Blitz}, L., {Fich}, M., \& {Stark}, A.~A. 1982, \apjs, 49, 183

\bibitem[{{Cash} {et~al.}(1980){Cash}, {Charles}, {Bowyer}, {Walter},
  {Garmire}, \& {Riegler}}]{Cash80}
{Cash}, W., {Charles}, P., {Bowyer}, S., {et~al.} 1980, \apjl, 238, L71

\bibitem[{{Finkbeiner}(2003)}]{Finkbeiner03}
{Finkbeiner}, D.~P. 2003, \apjs, 146, 407

\bibitem[{{F\"urst} {et~al.}(1990){F\"urst}, {Reich}, {Reich}, \&
  {Reif}}]{Fuerst90}
{F\"urst}, E., {Reich}, W., {Reich}, P., \& {Reif}, K. 1990, \aaps, 85, 691

\bibitem[{{Gao} {et~al.}(2011){Gao}, {Han}, {Reich}, {Reich}, {Sun}, \&
  {Xiao}}]{Gao11x}
{Gao}, X.~Y., {Han}, J.~L., {Reich}, W., {et~al.} 2011, \aap, 529, A159

\bibitem[{{Giardino} {et~al.}(2002){Giardino}, {Banday}, {G{\'o}rski},
  {Bennett}, {Jonas}, \& {Tauber}}]{Giardino02}
{Giardino}, G., {Banday}, A.~J., {G{\'o}rski}, K.~M., {et~al.} 2002, \aap, 387,
  82

\bibitem[{{Gold} {et~al.}(2011){Gold}, {Odegard}, {Weiland}, {Hill}, {Kogut},
  {Bennett}, {Hinshaw}, {Chen}, {Dunkley}, {Halpern}, {Jarosik}, {Komatsu},
  {Larson}, {Limon}, {Meyer}, {Nolta}, {Page}, {Smith}, {Spergel}, {Tucker},
  {Wollack}, \& {Wright}}]{Gold11}
{Gold}, B., {Odegard}, N., {Weiland}, J.~L., {et~al.} 2011, \apjs, 192, 15

\bibitem[{{Green}(2009)}]{Green09}
{Green}, D.~A. 2009, Bull. Astron. Soc. India, 37, 45

\bibitem[{{Haslam} {et~al.}(1982){Haslam}, {Salter}, {Stoffel}, \&
  {Wilson}}]{Haslam82}
{Haslam}, C.~G.~T., {Salter}, C.~J., {Stoffel}, H., \& {Wilson}, W.~E. 1982,
  \aaps, 47, 1

\bibitem[{{Haynes} {et~al.}(1978){Haynes}, {Caswell}, \& {Simons}}]{Haynes78}
{Haynes}, R.~F., {Caswell}, J.~L., \& {Simons}, L.~W.~J. 1978, Australian
  Journal of Physics Astrophysical Supplement, 45, 1

\bibitem[{{Hinshaw} {et~al.}(2007){Hinshaw}, {Nolta}, {Bennett}, {Bean},
  {Dor{\'e}}, {Greason}, {Halpern}, {Hill}, {Jarosik}, {Kogut}, {Komatsu},
  {Limon}, {Odegard}, {Meyer}, {Page}, {Peiris}, {Spergel}, {Tucker}, {Verde},
  {Weiland}, {Wollack}, \& {Wright}}]{Hinshaw07}
{Hinshaw}, G., {Nolta}, M.~R., {Bennett}, C.~L., {et~al.} 2007, \apjs, 170, 288

\bibitem[{{Howell} \& {Shakeshaft}(1966)}]{Howell66}
{Howell}, T.~F., \& {Shakeshaft}, J.~R. 1966, \nat, 210, 1318

\bibitem[{{Jarosik} {et~al.}(2011){Jarosik}, {Bennett}, {Dunkley}, {Gold},
  {Greason}, {Halpern}, {Hill}, {Hinshaw}, {Kogut}, {Komatsu}, {Larson},
  {Limon}, {Meyer}, {Nolta}, {Odegard}, {Page}, {Smith}, {Spergel}, {Tucker},
  {Weiland}, {Wollack}, \& {Wright}}]{Jarosik11}
{Jarosik}, N., {Bennett}, C.~L., {Dunkley}, J., {et~al.} 2011, \apjs, 192, 14

\bibitem[{{Kn{\"o}dlseder}(2000)}]{Knodlseder00}
{Kn{\"o}dlseder}, J. 2000, \aap, 360, 539

\bibitem[{{Kn{\"o}dlseder}(2004)}]{Knodlseder04}
{Kn{\"o}dlseder}, J. 2004 [arXiv:astro-ph/0407050] 

\bibitem[{{Kuchar} \& {Clark}(1997)}]{Kuchar97}
{Kuchar}, T.~A., \& {Clark}, F.~O. 1997, \apj, 488, 224

\bibitem[{{Landecker} {et~al.}(2010){Landecker}, {Reich}, {Reid}, {Reich},
  {Wolleben}, {Kothes}, {Uyan{\i}ker}, {Gray}, {Del Rizzo}, {F{\"u}rst},
  {Taylor}, \& {Wielebinski}}]{Landecker10}
{Landecker}, T.~L., {Reich}, W., {Reid}, R.~I., {et~al.} 2010, \aap, 520, A80

\bibitem[{{Miville-Desch{\^e}nes} \& {Lagache}(2005)}]{Miville05}
{Miville-Desch{\^e}nes}, M., \& {Lagache}, G. 2005, \apjs, 157, 302

\bibitem[{{Paladini} {et~al.}(2003){Paladini}, {Burigana}, {Davies}, {Maino},
  {Bersanelli}, {Cappellini}, {Platania}, \& {Smoot}}]{Paladini03}
{Paladini}, R., {Burigana}, C., {Davies}, R.~D., {et~al.} 2003, \aap, 397, 213

\bibitem[{{Paladini} {et~al.}(2005){Paladini}, {De Zotti}, {Davies}, \&
  {Giard}}]{Paladini05}
{Paladini}, R., {De Zotti}, G., {Davies}, R.~D., \& {Giard}, M. 2005, \mnras,
  360, 1545 (P05)

\bibitem[{{Pauliny-Toth} \& {Shakeshaft}(1962)}]{Toth62}
{Pauliny-Toth}, I.~K., \& {Shakeshaft}, J.~R. 1962, \mnras, 124, 61

\bibitem[{{Piddington} \& {Minnett}(1952)}]{Piddington52}
{Piddington}, J.~H., \& {Minnett}, H.~C. 1952, Australian Journal of Scientific
  Research A Physical Sciences, 5, 17

\bibitem[{{Reich} \& {Reich}(1988)}]{Reich88b}
{Reich}, P., \& {Reich}, W. 1988, \aaps, 74, 7

\bibitem[{{Reich} {et~al.}(1997){Reich}, {Reich}, \& {F{\"u}rst}}]{Reich97}
{Reich}, P., {Reich}, W., \& {F{\"u}rst}, E. 1997, \aaps, 126, 413

\bibitem[{{Reich} {et~al.}(2004){Reich}, {Reich}, \& {Testori}}]{PReich04}
{Reich}, P., {Reich}, W., \& {Testori}, J.~C. 2004, in The Magnetized
  Interstellar Medium, ed. B.~{Uyaniker}, W.~{Reich}, \& R.~{Wielebinski},
  63--68

\bibitem[{{Reich}(1982)}]{Reich82}
{Reich}, W. 1982, \aaps, 48, 219

\bibitem[{{Reich} {et~al.}(1990{\natexlab{a}}){Reich}, {F{\"u}rst}, {Reich}, \&
  {Reif}}]{Reich9011}
{Reich}, W., {F{\"u}rst}, E., {Reich}, P., \& {Reif}, K. 1990{\natexlab{a}},
  \aaps, 85, 633

\bibitem[{{Reich} {et~al.}(1990{\natexlab{b}}){Reich}, {Reich}, \&
  {F{\"u}rst}}]{Reich9021}
{Reich}, W., {Reich}, P., \& {F{\"u}rst}, E. 1990{\natexlab{b}}, \aaps, 83, 539

\bibitem[{{Reif} {et~al.}(1987){Reif}, {Reich}, {Steffen}, {M{\"u}ller}, \&
  {Weiland}}]{Reif87}
{Reif}, K., {Reich}, W., {Steffen}, P., {M{\"u}ller}, P., \& {Weiland}, H.
  1987, Mitteilungen der Astronomischen Gesellschaft Hamburg, 70, 419

\bibitem[{{Schneider} {et~al.}(2006){Schneider}, {Bontemps}, {Simon}, {Jakob},
  {Motte}, {Miller}, {Kramer}, \& {Stutzki}}]{Schneider06}
{Schneider}, N., {Bontemps}, S., {Simon}, R., {et~al.} 2006, \aap, 458, 855

\bibitem[{{Sharpless}(1959)}]{Sharpless59}
{Sharpless}, S. 1959, \apjs, 4, 257

\bibitem[{{Sofue} \& {Reich}(1979)}]{Sofue79}
{Sofue}, Y., \& {Reich}, W. 1979, \aaps, 38, 251

\bibitem[{{Sun} {et~al.}(2011{\natexlab{a}}){Sun}, {Reich}, {Reich}, {Xiao},
  {Gao}, \& {Han}}]{Sun11b}
{Sun}, X.~H., {Reich}, P., {Reich}, W., {et~al.} 2011{\natexlab{a}}, \aap, 536,
  A83

\bibitem[{{Sun} {et~al.}(2011{\natexlab{b}}){Sun}, {Reich}, {Han}, {Reich},
  {Wielebinski}, {Wang}, \& {M{\"u}ller}}]{Sun11a}
{Sun}, X.~H., {Reich}, W., {Han}, J.~L., {et~al.} 2011{\natexlab{b}}, \aap,
  527, A74 (S11)

\bibitem[{{Taylor} {et~al.}(1996){Taylor}, {Goss}, {Coleman}, {van Leeuwen}, \&
  {Wallace}}]{Taylor96}
{Taylor}, A.~R., {Goss}, W.~M., {Coleman}, P.~H., {van Leeuwen}, J., \&
  {Wallace}, B.~J. 1996, \apjs, 107, 239

\bibitem[{{Tweedy} {et~al.}(1995){Tweedy}, {Martos}, \&
  {Noriega-Crespo}}]{Tweedy95}
{Tweedy}, R.~W., {Martos}, M.~A., \& {Noriega-Crespo}, A. 1995, \apj, 447, 257

\bibitem[{{Wendker} {et~al.}(1991){Wendker}, {Higgs}, \&
  {Landecker}}]{Wendker91}
{Wendker}, H.~J., {Higgs}, L.~A., \& {Landecker}, T.~L. 1991, \aap, 241, 551

\bibitem[{{Xiao} {et~al.}(2011){Xiao}, {Han}, {Reich}, {Sun}, {Wielebinski},
  {Reich}, {Shi}, \& {Lochner}}]{Xiao11}
{Xiao}, L., {Han}, J.~L., {Reich}, W., {et~al.} 2011, \aap, 529, A15

\end{thebibliography}

\end{document}